\documentclass[a4paper,reqno,12pt,final]{amsart}
\usepackage{amsmath,amsfonts,amssymb,amsthm,amstext,eucal}
\usepackage{bbold}
\usepackage[latin1]{inputenc}
\DeclareMathOperator{\AdS}{AdS}
\DeclareMathOperator{\CP}{\CC P}
\DeclareMathOperator{\tr}{tr}
\newcommand{\drep}[1]{\underline{\boldsymbol{#1}}}
\renewcommand{\d}{\partial}
\newcommand{\1}{\mathbb{1}}
\newcommand{\Cl}{\mathrm{C}\ell}
\newcommand{\RR}{\mathbb{R}}

\newcommand{\CC}{\mathbb{C}}
\newcommand{\EE}{\mathbb{E}}
\newcommand{\SO}{\mathrm{SO}}
\newcommand{\CSO}{\mathrm{CSO}}
\newcommand{\ISO}{\mathrm{ISO}}

\newcommand{\fso}{\mathfrak{so}}
\newcommand{\fiso}{\mathfrak{iso}}
\newcommand{\fcso}{\mathfrak{cso}}
\newcommand{\Spin}{\mathrm{Spin}}
\newcommand{\eD}{\mathcal{D}}
\newcommand{\eS}{\mathcal{S}}
\newcommand{\fS}{\mathfrak{S}}
\newcommand{\fg}{\mathfrak{g}}

\newcommand{\fp}{\mathfrak{p}}
\newcommand{\fk}{\mathfrak{k}}
\newcommand{\half}{\tfrac{1}{2}}
\newcommand{\spb}{\$}
\begin{document}

\title[a maximally supersymmetric IIB background]{A new maximally
supersymmetric background of IIB superstring theory}
\author[Blau]{Matthias Blau}
\address{The Abdus Salam ICTP, Strada Costiera 11, 34014 Trieste,
Italy}
\email{mblau@ictp.trieste.it}
\author[Figueroa-O'Farrill]{José Figueroa-O'Farrill}
\address{Department of Mathematics and Statistics, The University of
Edinburgh, Edinburgh EH9 3JZ, UK}
\email{j.m.figueroa@ed.ac.uk}
\author[Hull]{Christopher Hull}
\address{Department of Physics, Queen Mary, University of London,
London E1 4NS, UK}
\email{c.m.hull@qmul.ac.uk}
\author[Papadopoulos]{George Papadopoulos}
\address{Department of Mathematics, King's College, Strand, London
WC2R 2LS, UK}
\email{gpapas@mth.kcl.ac.uk}
\thanks{EMPG-01-18, QMUL-PH-01-12}
\begin{abstract}
  We present a maximally supersymmetric IIB string background.  The
  geometry is that of a conformally flat lorentzian symmetric space
  $G/K$ with solvable $G$, with a homogeneous five-form flux.  We give
  the explicit supergravity solution, compute the isometries, the 32
  Killing spinors, and the symmetry superalgebra, and then discuss
  T-duality and the relation to M-theory.
\end{abstract}
\maketitle
\tableofcontents

\section{Introduction}

There are precisely four types of maximally supersymmetric solutions
of eleven-dimensional supergravity \cite{KG,Bonn,FOPmax}.  The first
three types are the familiar cases of flat eleven-dimensional space
(Minkowski space and its toroidal compactifications), $\AdS_4\times
S^7$ and $\AdS_7\times S^4$.  The fourth solution with 32 Killing
spinors is less well-known and was discovered by Kowalski-Glikman
\cite{KG}. In what follows we shall refer to this as the KG space or
solution.  Our purpose here is to show that IIB supergravity
\cite{SchwarzWestIIB,SchwarzIIB,HoweWestIIB} has, in addition to flat
space and $\AdS_5\times S^5$, another maximally supersymmetric
solution which is analogous to the KG space.

Eleven-dimensional supergravity, with bosonic fields the metric $ds^2$
and the four-form field strength $F_4$, has pp-wave solutions
\cite{Mwave} of the form
\begin{equation}
  \label{eq:metric}
  \begin{aligned}
  ds^2 &= 2 dx^+ dx^- + H(x^i,x^-) (dx^-)^2 + \sum_{i=1}^9 (dx^i)^2\\
  F_4&= dx^-\wedge \varphi
 \end{aligned}
\end{equation}
where
$H(x^i,x^-)$ obeys
\begin{equation}
  \label{eq:poisson}
  \bigtriangleup H= \frac {1}{12} |\varphi|^2 .
\end{equation}
Here $\bigtriangleup$ is the laplacian in the transverse euclidean
space $\EE^9$ with coordinates $x^i$ and $\varphi(x^i,x^-)$ is (for
each $x^-$) a closed and coclosed $3$-form in $\EE^9$.  Observe that
$\partial/\partial x^+$ is a covariantly constant null vector.  This
solution preserves at least 16 supersymmetries.

An interesting subclass of these metrics are those in which
\begin{equation}
  \label{eq:cwtype}
  H(x^i,x^-)= \sum_{i,j} A_{ij} x^i x^j
\end{equation}
where $A_{ij} = A_{ji}$ is a constant symmetric matrix. Remarkably,
these metrics describe lorentzian symmetric spaces $G/K$ where
$K=\RR^9$ and $G$ is a solvable group depending on $A$
\cite{CahenWallach,FOPflux}.  Moreover, they are indecomposable (that
is, they are not locally isometric to a product) when the matrix
$A_{ij}$ is non-degenerate \cite{CahenWallach}.  As $H$ grows
quadratically with $|x^i|$, these are not conventional gravitational
wave solutions.  Such solutions with covariantly constant $F_4$ are
spacetimes with a null homogeneous flux, referred to as Hpp-waves in
\cite{FOPflux}.  The solutions are parameterised (up to overall scale
and permutations) by the eigenvalues of $A_{ij}$ and all have at least
16 Killing spinors.  Surprisingly, there is precisely one non-trivial
choice of $A_{ij}$ (up to diffeomorphisms and overall scale
transformations), for which the solution has 32 Killing spinors.  This
is the KG solution, with
\begin{equation}
  \label{eq:Aij}
  \begin{aligned}[m]
    A_{ij}& =
    \begin{cases}
      -\tfrac19 \mu^2 \delta_{ij} & i,j=1,2,3\\
      -\tfrac1{36} \mu^2 \delta_{ij} & i,j=4,5,\dots,9
    \end{cases}\\
    \varphi &=\mu dx^1\wedge dx^2\wedge dx^3~,
  \end{aligned}
\end{equation}
where $\mu$ is a parameter which can be set to any nonzero value by a
change of coordinates.

The IIB supergravity also has pp-wave solutions with metric of the
form \eqref{eq:metric} but now the transverse space is $\EE^8$ and so
$i=1,...,8$.  In particular, we shall show that if the only
non-vanishing flux is that of the self-dual five-form field strength
$F_5$ and the dilaton is constant, then there is a pp-wave solution
with metric similar to that of \eqref{eq:metric}. The self-dual
five-form $F_5$ is null and associated with a closed self-dual 4-form
$\varphi$ on the transverse $\EE^8$ provided that $H$ satisfies an
equation similar to \eqref{eq:poisson}.  Again these solutions
generically preserve 16 supersymmetries. If $H$ is of the form
\eqref{eq:cwtype}, then the metric again describes a lorentzian
symmetric space $G/\RR^8$; the solvable group $G$ will be discussed in
detail in Section~\ref{sec:symmetry}.  This solution describes an
Hpp-wave with null homogeneous five-form flux.  Furthermore, we will
show that there is precisely one choice of $A_{ij}$ (up to
diffeomorphisms and overall scale transformations) for which the
solution has 32 Killing spinors.  In particular, we shall find that
\begin{equation}
  \label{eq:maxsol}
  \begin{aligned}[m]
    ds^2 &= 2 dx^- dx^+ - 4 \lambda^2 \sum_{i=1}^8 (x^i)^2 (dx^-)^2 +
    \sum_{i=1}^8 (dx^i)^2\\
    F_5 &= \lambda dx^- \wedge \left(dx^1\wedge dx^2\wedge dx^3 \wedge
      dx^4 + dx^5\wedge dx^6\wedge dx^7 \wedge dx^7\right)~
  \end{aligned}
\end{equation}
preserves all supersymmetry of IIB supergravity, where $\lambda$ is a
parameter which can be set to any desired nonzero value by a
coordinate transformation.  For this solution
\begin{equation}
  \label{eq:Bij}
  \begin{aligned}[m]
    A_{ij}& = -4\lambda^2 \delta_{ij} \\
    \varphi &=\lambda ( \omega + \star\omega)~,
  \end{aligned}
\end{equation}
where
\begin{equation}
  \label{eq:4form}
  \omega =dx^1\wedge dx^2\wedge dx^3 \wedge dx^4
\end{equation}
and $\star$ denotes the Hodge dual in the transverse $\EE^8$.

The isometry groups of the 11-dimensional solutions $\AdS_4\times S^7$
and $\AdS_7\times S^4$ are $\SO(3,2)\times\SO(8)$ and $\SO(6,2)\times
\SO(5)$, respectively, both $38$-dimensional.  The isometry group of
the KG solution \cite{CKG,FOPflux} is again $38$-dimensional but it is
non-semisimple.  In fact, it is isomorphic to the extension, by a
one-dimensional group of outer automorphisms, of the semidirect
product $H(9)\rtimes (\SO(3)\times \SO(6))$, with $H(9)$ a
$9$-dimensional Heisenberg group.  This semidirect product group is
the quotient by a central element of $\CSO(3,2) \times \CSO(6,2)$, a
certain group contraction\footnote{The group $\CSO(p,q)$ was defined
  in \cite{HullGauged} as the group contraction of $\SO(p+q)$ or
  $\SO(p,q)$ preserving a metric with $p$ positive eigenvalues and $q$
  zero ones, so that, for example, $\CSO(p,1)$ is the Euclidean group
  $\ISO(p)$.} of both $\SO(3,2)\times\SO(8)$ and of
$\SO(6,2)\times\SO(5)$.  For the IIB theory, the isometry group of
$\AdS_5\times S^5$ is the $30$-dimensional group $\SO(4,2)\times
\SO(6)$, while the isometry group of our new maximally supersymmetric
space is again $30$-dimensional and, as we will show, is related to
the $\CSO(4,2) \times \CSO(4,2)$ contraction of $\SO(4,2)\times
\SO(6)$.

The physical significance of these solutions in string and M-theory is
puzzling, particularly for the Hpp-wave solutions that preserve all
supersymmetry which are in the same universality class of solutions as
Minkowski and $\AdS\times S$ spaces. Nevertheless, unlike the M-theory
Hpp-wave, the IIB Hpp-wave that we have found in this paper may be
more amenable to investigation at least in string perturbation theory.
This is because the dilaton in this background is constant and so the
string coupling constant can be kept small everywhere.  Arguments
similar to those in \cite{KaRa} which utilise the form of the geometry
and the 32 supersymmetries should imply that the new solution is an
exact solution of IIB string theory and that the KG solution should be
an exact solution of M-theory. However, such arguments are rather
formal in the absence of an explicit quantisation of string theory in
RR backgrounds or of M-theory.

\section{IIB field equations and Killing spinors }

We shall consider IIB supergravity backgrounds specified by a triple
$(M,g,F_5)$ where $M$ is a ten-dimensional oriented spin manifold, $g$
is a lorentzian metric and $F_5$ is a self-dual five-form. The dilaton
is constant and all other fields are set to zero.  The relevant field
equations are the Einstein equations and the self-duality constraint
on $F_5$.  A Killing spinor for such a background need only satisfy
the equation associated with the vanishing of the gravitino
supersymmetry transformation, as all other conditions are satisfied
automatically for our ansatz. For the maximally supersymmetric
solution, the Killing spinor equations imply the field equations.

In our conventions, the metric is mostly plus with signature $(9,1)$.
The Clifford algebra $\Cl(9,1)$, into which the spin group
$\Spin(9,1)$ embeds, admits a basis spanned by real
gamma-matrices $\Gamma^a$ satisfying
\begin{equation*}
  \left\{ \Gamma^a, \Gamma^b \right\} = + 2 \eta^{ab}\, \1
\end{equation*}
such that $\Gamma^0$ is skewsymmetric and hence skewhermitian, whereas
the $\Gamma^i$, for $i=1,\dots,9$ are symmetric and hence hermitian.
In this basis, the charge conjugation matrix $C$ is $\Gamma^0$.  The
group $\Spin(9,1)$ has two real irreducible representations of
dimension $16$: $S_+$ and $ S_-$, the spinors of positive and negative
chirality, respectively.  Let $\spb_\pm$ denote the corresponding spin
bundles over $M$.  The two spinor parameters in the supersymmetry
transformations have positive chirality, so each is a section of
$\spb_+$.  It is convenient to combine the two spinors into a single
complex-valued 16-component chiral spinor field, which is a section of
the complexified Weyl spinor bundle $\spb := \spb_+ \otimes \CC$.

The Einstein equation for our ansatz is (in the Einstein frame)
\begin{equation*}
  R_{MN} - \tfrac12 g_{MN} R= \tfrac16 F_{L_1\cdots L_4M} F^{L_1\cdots
  L_4}{}_{N}~,
\end{equation*}
where $M,N=0,1,\dots,9$. In the string frame, the background $(M,g,F)$
is invariant under an infinitesimal supersymmetry transformation with
spinor parameter $\varepsilon$ if and only if $\varepsilon$ satisfies
the Killing spinor equation
\begin{equation*}
  \eD_M\varepsilon=0\end{equation*}
where the supercovariant derivative is
\begin{equation*}
 \label{eq:kse}
  \eD_M \varepsilon = \nabla_M \varepsilon + \tfrac{i}{192} e^{\phi}
  F_{ML_1\cdots L_4} \Gamma^{L_1 \cdots L_4} \varepsilon~.
\end{equation*}
and $\phi$ is the (constant) dilaton, $\nabla$ is the spin connection
and products $\Gamma_{M...N}$ of $\Gamma$ matrices are skewsymmetrised
with strength one.  It is convenient to rescale the $5$-form to absorb
the prefactor, bringing the connection to the simpler form
\begin{equation}
  \label{eq:superconn}
  \eD_M \varepsilon = \nabla_M \varepsilon + \tfrac{i}{24}
  F_{ML_1\cdots L_4} {\Gamma^{L_1 \cdots L_4}  \varepsilon}~.
\end{equation}
In what follows we will present a background $(M,g,F)$ for which this
connection is flat: $(M,g)$ will be an indecomposable lorentzian
symmetric space $G/ K$ with solvable $G$ (referred to as a
Cahen--Wallach (CW) space in \cite{FOPflux}) and $F$ a parallel null
form.  This background is the IIB analogue of a maximally
supersymmetric solution of M-theory originally discovered in
\cite{KG}.  It is satisfying to see that in this way it is not just
the $\AdS \times S$ vacua that M-theory and IIB string theory have in
common, but also these more exotic ones.

\section{A family of lorentzian symmetric spaces}

The classification of riemannian symmetric spaces is a classic piece
of mathematics dating back to the work of Élie Cartan.  In contrast,
the classification of pseudo-riemannian spaces is a much harder
problem, because there is no decomposition theorem which reduces the
problem to studying symmetric splits of a Lie algebra $\fg = \fk
\oplus \fp$ such that the natural action of $\fk$ on $\fp$ is
irreducible. In fact, one is unavoidably led to consider those cases
where $\fk$ acts reducibly but indecomposably.  The canonical example
is a space which possesses a parallel null vector, since the existence
of such a vector reduces the holonomy, but does not necessarily
decompose the space, as the metric is degenerate when restricted to
the null direction.  Despite this difficulty, Cahen and Wallach
\cite{CahenWallach} classified the lorentzian symmetric spaces;
although so far the general pseudo-riemannian case remains
unclassified.  Cahen and Wallach proved that in dimensions $d\geq 3$
an indecomposable lorentzian symmetric space is locally isometric
either to (anti-) de~Sitter space or to a metric in a
$(d-3)$-dimensional family of symmetric spaces having a solvable
transvection group.\footnote{If a symmetric space is of the form
  $G/K$, then $G$ is called the transvection group.}  It is precisely
this family of spaces which plays a role in the construction described
in this paper.  A similar construction was discussed recently in
\cite{FOPflux} in the context of eleven-dimensional supergravity. The
same construction applies here, and we refer the reader to that paper
for more details.

Consider the metric
\begin{equation}
  \label{eq:cwmetric}
  ds^2 = 2 dx^+ dx^- + \sum_{i,j=1}^8 A_{ij} x^i x^j (dx^-)^2 +
  \sum_{i=1}^8 dx^i dx^i~,
\end{equation}
where $A_{ij} = A_{ji}$ is a symmetric matrix and $i=1,\dots,8$.  This
metric is indecomposable if and only if $A_{ij}$ is nondegenerate.  If
$A_{ij}\not\equiv 0$, we can rotate the coordinates $x^i$ and rescale $x^\pm
\to c^{\pm 1} x^\pm$ to bring $A_{ij}$ to a diagonal form with
eigenvalues lying on the unit sphere in $\RR^8$.  The moduli space of
the indecomposable CW metrics \eqref{eq:cwmetric} is therefore
seven-dimensional: the seven-sphere minus the singular locus
consisting of degenerate $A_{ij}$'s and modded out by the natural
action of the symmetric group $\fS_8$.

We introduce a coframe $\theta^a = (\theta^{\widehat{\imath}},
\theta^{\widehat{+}}, \theta^{\widehat{-}})$, given by
\begin{equation}
  \label{eq:coframe}
  \theta^{\widehat{\imath}} = dx^i \qquad \theta^{\widehat{-}} = dx^- \qquad
\theta^{\widehat{+}} = dx^+ +
  \half \sum_{i,j} A_{ij} x^i x^j dx^-~,
\end{equation}
so that the metric is
\begin{equation*}
  ds^2 = 2 \theta^{\widehat{+}}\theta^{\widehat{-}} +
  \sum_{\widehat{\imath}=1}^8 \theta^{\widehat{\imath}}
  \theta^{\widehat{\imath}}~.
\end{equation*}
Frame indices are raised and lowered using the flat metric
$\eta_{ab}$.  Note that we do not need to distinguish between the
upper indices $\widehat{\imath},\widehat{-}$ and $i,-$ or between the
lower indices $\widehat{\imath},\widehat{+}$ and $i,+$ , and in what
follows we will omit the hats except for upper flat indices
$\widehat{+}$ and lower ones $\widehat{-}$.

The first structure equation
\begin{equation*}
  d\theta^a + \omega^a{}_b \wedge \theta^b = 0~,
\end{equation*}
gives the only nonzero components of the spin-connection $\omega^{ab}$
as
\begin{equation}
  \label{eq:conn1form}
  \omega^{\widehat{+}i} = -\omega^{i\widehat{+}} = \sum_j A_{ij} x^j
  dx^-~.
\end{equation}
The covariant derivative on spinors is given by
\begin{equation*}
  \nabla_M = \d_M + \half \omega_M{}^{ab} \Sigma_{ab}~,
\end{equation*}
where $\Sigma_{ab} = \half \Gamma_{ab}$ are the spin generators.
In our case we see that $\nabla_+ = \d_+$ and $\nabla_i = \d_i$,
whereas
\begin{equation*}
  \nabla_- = \d_- + \half \sum_{i,j} A_{ij} x^j \Gamma_+ \Gamma_i~.
\end{equation*}

The components of the Riemann curvature tensor of $(M,g)$ are
given by
\begin{equation*}
  R_{-i-j} = - A_{ij}~,
\end{equation*}
with all other components not related to this by symmetry vanishing.
The Ricci curvature has nonzero component $R_{--} = -\tr A$, and the
scalar curvature vanishes.  In particular, $(M,g)$ is Ricci-flat if
and only if $A_{ij}$ is trace-free.  It also follows that $(M,g)$ is
conformally flat if and only if $A_{ij}$ is a scalar matrix; that is,
$A_{ij} \propto \delta_{ij}$.

The holonomy of the metric \eqref{eq:cwmetric} is $\RR^8 \subset
\SO(9,1)$.  In terms of Lorentz transformations, the holonomy group
acts by null rotations.  The parallel forms are the constant
functions, together with $dx^- \wedge \varphi$ where $\varphi$ is any
constant-coefficient form
\begin{equation*}
  \varphi = \sum_{1\leq i_1<i_2<\cdots<i_p \leq 8}~ c_{i_1i_2\cdots
  i_p} dx^{i_1} \wedge dx^{i_2} \wedge \cdots \wedge dx^{i_p}~.
\end{equation*}
Notice, in particular, that all such parallel forms are null.

\section{A maximally supersymmetric solution and Killing spinors}

Motivated by the KG solution, we consider the ansatz
\begin{equation}
  \begin{aligned}
    \label{eq:ansatz}
    ds^2 &= 2 dx^+ dx^- + \sum_{i,j=1}^8 A_{ij} x^i x^j (dx^-)^2 +
    \sum_{i=1}^8 dx^i dx^i \\
    F_5 &= \lambda \, dx^- \wedge (dx^1 \wedge dx^2 \wedge dx^3 \wedge
    dx^4 + dx^5 \wedge dx^6 \wedge dx^7 \wedge dx^8)~,
  \end{aligned}
\end{equation}
where $\lambda$ is a real constant.  The metric is clearly that of a
CW space and $F_5$ is chosen to be null and self-dual.  It was seen in
the previous section that $F_5$ is parallel.

In order to write the supersymmetric covariant derivative
\eqref{eq:kse} in the above ansatz, let us decompose it as $\eD_M =
\nabla_M + \Omega_M$, where
\begin{equation}
  \label{eq:Omega}
  \Omega_M :=
  \begin{cases}
    0 & M=+\\
    \lambda ( I + J ) & M=-\\
    - \lambda \Gamma_+ \Gamma_M I & M=1,2,3,4\\
    - \lambda \Gamma_+ \Gamma_M J & M=5,6,7,8
  \end{cases}
\end{equation}
and we have introduced the notation $I:=\Gamma_{1234}$ and
$J:=\Gamma_{5678}$.   Notice
that $I^2 = J^2 = \1$ and $IJ = JI$.  Finally notice that since
$\Gamma_+^2=0$,
\begin{equation}
  \label{eq:omegaomega}
  \Omega_i \Omega_j = 0 \qquad\text{for all $i,j=1,2,\dots,8$}~.
\end{equation}

We now determine the precise form of the metric (i.e., the matrix
$A_{ij}$) for which the supersymmetric covariant derivative
\eqref{eq:superconn} is flat, this being a necessary (and locally
sufficient) condition for maximal supersymmetry.

Let $\varepsilon$ be a Killing spinor; that is, a section of $\spb$
obeying $\eD \varepsilon = 0$.  Since $\nabla_+ = \d_+$ and $\Omega_+
= 0$, we see that $\varepsilon$ is independent of $x^+$.  Similarly
from
\begin{equation}
  \label{eq:killi}
  \d_j \varepsilon = - i \Omega_j \varepsilon
\end{equation}
and equation \eqref{eq:omegaomega}, we see that $\d_i \d_j \varepsilon
= 0$, whence $\varepsilon$ is at most linear in $x^i$.  Let us write
it as
\begin{equation*}
  \varepsilon = \chi + \sum_j x^j \varepsilon_j~,
\end{equation*}
where the spinors $\chi$ and $\varepsilon_i$ are only
functions of $x^-$.  From equation \eqref{eq:killi} we see that
$\varepsilon_j = - i \Omega_j \chi$, so that  any Killing spinor
$\varepsilon$ takes the form
\begin{equation}
  \label{eq:kill0}
  \varepsilon = \left(\1 - i \sum_j x^j \Omega_j \right)
  \chi~,
\end{equation}
where the spinor $\chi$ only depends on $x^-$.  The dependence on
$x^-$ is fixed from the one remaining equation
\begin{equation*}
  \d_- \varepsilon + \half \sum_{i,j} A_{ij} x^j \Gamma_+ \Gamma_i
  \varepsilon + i \lambda (I + J) \varepsilon = 0~,
\end{equation*}
which will also imply an integrability condition fixing the matrix
$A_{ij}$.

Inserting the above expression \eqref{eq:kill0} for $\varepsilon$ into
this equation and after a little bit of algebra (using repeatedly that
$\Gamma_+^2 = 0$), we find
\begin{multline*}
  \chi' + i \lambda (I+J) \chi\\
  + \sum_i x^i \left( \half \sum_j A_{ij} \Gamma_+ \Gamma_j + \lambda
    (I+J) \Omega_i - \lambda \Omega_i (I+J) \right)\, \chi = 0~,
\end{multline*}
where ${}'$ denotes derivative with respect to $x^-$.
 Using
\begin{equation*}
  (I+J) \Omega_i - \Omega_i (I+J) = 2 \lambda \Gamma_+ \Gamma_i
  \qquad\text{for all $i$}~,
\end{equation*}
this can be rewritten as
\begin{equation*}
  \chi' + i \lambda (I+J) \chi
  = \sum_i x^i \left(
 2\lambda^2  \Gamma_i + \half \sum_j A_{ij}
  \Gamma_j
 \right)\, \Gamma_+ \chi ,
\end{equation*}

The left-hand side is independent of $x^i$ as $\chi$ is, while the
right-hand side has explicit $x^i$ dependence, and so both sides must
vanish separately.  The vanishing of the left-hand side is a
first-order linear ordinary differential equation with constant
coefficients, which has a unique solution for each initial value.  The
number of supersymmetries is then the dimension of the space of such
initial values for which the right-hand side vanishes.  The right-hand
side will vanish for any initial value spinor that is annihilated by
$\Gamma_+$, and so there are always (at least) 16 real (or 8 complex)
Killing spinors, and the solution is half-BPS for arbitrary $A_{ij}$.
However, if $A_{ij}$ is chosen so that the bracket on the right-hand
side vanishes, then the right-hand side vanishes for arbitrary spinors
$\chi$, and the solution is maximally supersymmetric.

The right-hand side will vanish   for all $\chi$ if and only if
\begin{equation*}
  A_{ij} = - 4\lambda^2 \delta_{ij}~,
\end{equation*}
which, with ansatz \eqref{eq:ansatz}, becomes the solution
\eqref{eq:maxsol} presented in the introduction.  The parameter
$\lambda$, which hides the dependence on the (constant) dilaton, can
be set to any desired (nonzero) value by rescaling the coordinates
$x^\pm\mapsto c^{\pm 1} x^\pm$.  We remark that for this choice of
$A_{ij}$, the CW metric \eqref{eq:cwmetric} is conformally flat.

The Killing spinors are easy to find, as they obey a first order
ordinary differential equation
\begin{equation*}
  \chi' = - i \lambda (I + J) \chi~.
\end{equation*}
The solution is given by the matrix exponential
\begin{equation*}
  \chi(x^-) = \exp\left(- \lambda x^- i (I + J) \right)
  \psi,
\end{equation*}
where $\psi = \chi(0)$ is an arbitrary constant spinor.  Since $I$ and
$J$ commute, we can write the solution as
\begin{equation*}
  \chi(x^-) = \exp\left(- i \lambda x^- I\right) \exp\left(- i \lambda
  x^- J \right) \psi~,
\end{equation*}
and since $I^2 = J^2 = \1$, we can write this as
\begin{equation*}
  \chi(x^-) = \left(\cos(\lambda x^-) \1 - i \sin(\lambda x^-)  I
  \right) \left(\cos(\lambda x^-) \1 - i \sin(\lambda x^-) J \right)
  \psi~,
\end{equation*}
from where the Killing spinors $\varepsilon$ can be read off using
\eqref{eq:kill0}. Indeed we find that
\begin{multline}
  \label{eq:iibpp}
    \varepsilon(\psi)=\left(1-i\sum_{j=1}^8
      x^j\Omega_j\right) \left(\cos(\lambda x^-) \1 - i \sin(\lambda
      x^-) I \right)\\
    {}\times  \left(\cos(\lambda x^-) \1 - i \sin(\lambda x^-) J \right)
    \psi~.
\end{multline}

The Killing spinor depends non-trivially on all the coordinates of
spacetime apart from $x^+$. It is periodic in $x^-$ with period
$2\pi/\lambda$.

\section{A IIB pp-wave in a homogeneous flux }

IIB supergravity admits a more general solution than those that have
been discussed so far.  In particular,
\begin{equation}
  \label{eq:nmaxsol}
  \begin{aligned}[m]
    ds^2 &= 2 dx^+ dx^- + H(x^i,x^-) (dx^-)^2 +
    ds^2(\EE^8)~\\
    F_5&= dx^-\wedge (\omega+ \star \omega)
  \end{aligned}
\end{equation}
is a solution provided that for fixed $x^-$, $H:\EE^8 \to \RR$
satisfies the Poisson equation
\begin{equation}
  \bigtriangleup H=-32 \vert \omega \vert ^2
\end{equation}
where $\omega(x^i,x^-)$ is, for each, $x^-$, a closed and co-closed
$4$-form in the transverse space $\EE^8$.  For example if
$\omega=\lambda dx^1\wedge dx^2\wedge dx^3\wedge dx^4$, one can set
\begin{equation}
  H= \frac{q}{|x|^6} + A_{ij} x^ix^j~,
\end{equation}
where
\begin{equation}
  \tr  A = - 32 \lambda^2~.
\end{equation}
Such a solution has the interpretation of a IIB pp-wave in the
presence of null homogeneous $F_5$ flux.  Generic such solutions have
16 Killing spinors. These are
\begin{equation*}
  \varepsilon(\psi)=
  \left(\cos(\lambda x^-) \1 - i \sin(\lambda x^-) I \right)
  \left(\cos(\lambda x^-) \1 - i \sin(\lambda x^-) J \right)
  \psi~,
\end{equation*}
where now the constant spinor $\psi$ satisfies, $\Gamma_+\psi=0$.
Again the Killing spinors are independent of $x^+$ and periodic in the
null coordinate $x^- $ with period $2\pi/\lambda$, but they are now
also independent of the $x^i$.

\section{The symmetry superalgebra}
\label{sec:symmetry}

The maximally supersymmetric solution \eqref{eq:maxsol} is invariant
under the Lie algebra $\fS_0$ generated by those Killing vectors which
also preserve $F_5$, and under the supersymmetries generated by the
$32$ Killing spinors. Together these symmetries form a Lie
superalgebra $\fS$ whose even part is $\fS_0$ and whose odd part
$\fS_1$ is $32$-dimensional.  The structure of $\fS$ can be found by
calculating the symmetry algebra \cite{JMFKilling,PKT}, as was done
for the KG solution in \cite{FOPflux}.

The isometries of the CW space $(M,g)$ were discussed in detail in
\cite{FOPflux}.  The isometry algebra of the metric $ds^2$ consists of
two parts.  The first part is the $18$-dimensional Lie algebra $\fg$
of the transvection group $G$, with generators we denote
$\{e_i^*,e_i,e_+,e_-\}$; whereas the second part is the subalgebra of
$\fso(8)$ leaving $A_{ij}$ invariant.  In this case, since $A_{ij}$ is
diagonal, this is all of $\fso(8)$, with generators $M_{ij}$ and
corresponding Killing vectors $\xi_{M_{ij}} = x^i\partial_j -
x^j\partial_i$.  However not all these isometries are symmetries of
the full solution because $F_5$ is not invariant under all of
$\fso(8)$, but only under a subalgebra $\fso(4) \oplus \fso(4) \subset
\fso(8)$.  In summary, the complete set of Killing vectors preserving
the supergravity solution \eqref{eq:maxsol} is the following:
\begin{equation*}
  \begin{aligned}[m]
    \xi_{e_+}&= -\d_+, \qquad\qquad  \xi_{e_-}=-\d_-~,\\
    \xi_{e_i}&=- \cos(2\lambda x^-) \d_i - 2\lambda \sin(2\lambda x^-)
    x^i \d_+\quad i=1,\dots,8,\\
    \xi_{e^*_i}&= - 2\lambda \sin(2\lambda x^-) \d_i + 4\lambda^2
    \cos(2\lambda x^-) x^i \d_+~ \quad i=1,\dots,8,\\
    \xi_{M_{ij}}&=x^i\partial_j-x^j\partial_i \qquad i,j=1,\dots,4~
    {\rm and}~ i,j=5,\dots,8~,
  \end{aligned}
\end{equation*}
where we denote the Killing vector field associated with $X\in\fS_0$
by $\xi_X$.

As $F_5$ is parallel, it is homogeneous; that is, invariant under the
transvection group.  It is straightforward to calculate the Lie
algebra, finding $\fS_0=\fg \rtimes (\fso(4) \oplus \fso(4))$. If
$\{M_{ij}\}$, for $i,j=1,\dots,4$ and $i,j=5,\dots,8$ are the
generators of $\fso(4) \oplus \fso(4)$, then the Lie brackets of
$\fS_0$ are
\begin{equation}
  \label{eq:S0S0}
  \begin{aligned}[m]
    [e_-,e_i] &= e^*_i \qquad [e_-,e^*_i] = - 4\lambda^2 e_i\qquad
    [e^*_i,e_j]= -4\lambda^2 \delta_{ij} e_+\\
    [M_{ij}, e_k]&= -\delta_{ik} e_j+\delta_{jk} e_i \qquad [M_{ij},
    e^*_k]= -\delta_{ik} e^*_j+\delta_{jk} e^*_i~,
  \end{aligned}
\end{equation}
where, in the second line, $1\leq i,j,k\leq 4$ and $5\leq i,j,k\leq
8$.  The symmetry algebra $\fS_0$ of the solution \eqref{eq:maxsol} is
$30$-dimensional and non-semisimple.  It is interesting to note that
the dimension of this algebra coincides with that of the $\AdS_5
\times S^5$ maximally supersymmetric solution.  A similar fact has
previously been observed for eleven-dimensional supergravity and we
believe this ``coincidence'' deserves further investigation.

The spacetime is a coset space $G/K$ where $K$ is the $\RR^8$ subgroup
of $G$ whose Lie algebra is generated by $\{e_i^*\}$.  Equivalently,
we may think of the spacetime as a connected component of the coset
$\eS /(\ISO(4)\times \ISO(4))$, where $\eS$ the isometry group with
Lie algebra $\fS_0$ and where $\ISO(4)$ is the Euclidean group $\RR^4
\rtimes \SO(4)$.  The Lie algebra of $\ISO(4)\times\ISO(4)$ is
generated by $\{e_i^*,M_{ij}\}$.

To explore the structure of the symmetry algebra $\fS_0 = \fg \rtimes
(\fso(4) \oplus \fso(4))$ further, it will be useful to decompose the
$\fso(8)$ vectors $v^i$ into $\fso(4)\oplus \fso(4)$ representations,
$v^a,v^{a'}$, where $a,b=1,...,4$ are vector indices for the first
$\fso(4)$, and $a',b'=1,...,4$ are vector indices for the second
$\fso(4)$.  Then $\{M_{ab},e_a\}$ and $\{M_{ab},e^*_a\}$ both generate
$\fiso(4)$ subalgebras.  The subalgebra of $\fS_0$ with generators
$\{M_{ab},e_a ,e^*_a,e_+\}$ is isomorphic to the Lie algebra
$\fcso(4,1,1)$, as defined in \cite{HullGauged}.  This arises as a
contraction of $\fso(5,1)$ as follows.  Decomposing the 15 generators
into $\fso(4)\oplus \fso(1,1)$ representations, one obtains the
$\fso(4)$ generators $\{M_{ab}\}$, the $\fso(1,1)$ generator $U$ and
the generators $Y_{a,i}$ in the $(\drep{4},\drep{2})$ representation.
Rescaling the $Y$ by a factor $\xi$ and $U$ by a factor $\xi^2$ and
then taking the limit $\xi \to 0$ gives the contracted Lie algebra of
$\fcso(4,1,1)$, with the algebra given as above with $U=e_+$ and
$(Y_{ai})=(e_a, e_a^*)$.
  
Similarly, there is a $\fcso(4,1,1)$ generated by $\{M_{a'b'},e_{a'}
,e^*_{a'}, e_+\}$.  The algebra generated by $\{M_{ij},e_i,e^*_i,
e_+\}$ is closely related to $\fcso(4,2)\oplus \fcso(4,2)$.  Indeed if
we let $\{M_{ab},e_a ,e^*_a,U\}$ denote the generators of the first
$\fcso(4,1,1)$ and we let $\{M_{a'b'},e_{a'} ,e^*_{a'},U'\}$ be those
of the second, then the algebra generated by $\{M_{ij},e_i,e^*_i,
e_-\}$ is obtained from $\fcso(4,1,1)\oplus \fcso(4,1,1)$ by setting
$U=U'=e_+$, so that the Lie algebra is the quotient
$\big(\fcso(4,1,1)\oplus \fcso(4,1,1)\big)/\RR$ by the central
subalgebra generated by $U-U'$.  Finally, we extend by the outer
automorphism generated by $e_-$, and we see that $\fS_0$ is an
extension of $\big(\fcso(4,1,1)\oplus \fcso(4,1,1)\big)/\RR$ by the
single generator $e_-$.
  
If one were to periodically identify the $x^+$ coordinate, then the
non-compact generator $e_+$ becomes a compact generator of
translations in this circular direction, and $\fcso(4,1,1)$ in the
above becomes $\fcso(4,2)$, which is the contraction of $\fso(6)$ or
$\fso(4,2)$ obtained by decomposing into $\fso(4)\oplus\fso(2)$
representations and scaling as above.  Then $\fS_0$ becomes an
extension of $\big(\fcso(4,2)\oplus \fcso(4,2)\big)/\fso(2)$ by the
single generator $e_-$, which will be a compact generator if $x^-$ is
periodically identified, and non-compact otherwise.  Similarly, the
symmetry algebra for the KG solution with periodic $x^+$ is an
extension of $\big(\fcso(3,2)\oplus \fcso(6,2)\big)/\fso(2)$ by a
single generator, while for non-periodic $x^+$ it is an extension of
$\big(\fcso(3,1,1)\oplus \fcso(6,1,1)\big)/\RR$.  Other real forms
arise if one compactifies a coordinate $z=ax^++bx^-$.

To compute the commutator of the even and odd generators of the
symmetry superalgebra, we also need to find the spinorial Lie
derivative of the Killing spinors along the Killing vector directions.
The spinorial Lie derivative $L_\xi$ along a Killing vector direction
$\xi$ is defined as
\begin{equation}
  L_\xi = \nabla_\xi + \tfrac14 \nabla_M \xi_N \Gamma^{MN}~.
\end{equation}
Properties of the spinorial Lie derivative have been given in
\cite{JMFKilling,FOPflux}.  For our solution \eqref{eq:maxsol}, the
non-vanishing spinorial Lie derivatives of the Killing spinors are
\begin{equation*}
  \begin{aligned}[m]
    L_{e_-}\varepsilon(\psi)&=\varepsilon(i\lambda (I+J) \psi)\\
    L_{e_i}\varepsilon(\psi)&=\varepsilon(-i\lambda I \Gamma^i
    \Gamma_+\psi) \quad i=1,\dots,4\\
    L_{e_i}\varepsilon(\psi)&=\varepsilon(-i\lambda J \Gamma^i
    \Gamma_+\psi) \quad i=5,\dots,8\\
    L_{e^*_i}\varepsilon(\psi)&=\varepsilon(-2\lambda^2 I \Gamma^i
    \Gamma_+\psi) \quad i=1,\dots,4\\
    L_{e^*_i}\varepsilon(\psi)&=\varepsilon(-2\lambda^2 J \Gamma^i
    \Gamma_+\psi) \quad i=5,\dots,8\\
    L_{M_{ij}}\varepsilon(\psi)&=\varepsilon(\half \Gamma^{ij}\psi)
    \quad i,j=1,\dots,4~ \text{and}~ i,j=5,\dots,8~,
  \end{aligned}
\end{equation*}
where $L_X$ for $X\in\fg$ stands for $L_{\xi_X}$. Using these
expressions, the commutators of the even and odd generators of the
symmetry superalgebra are
\begin{equation*}
  \begin{aligned}[m]
    [e_+, Q]&=0\qquad [e_-, Q]=i\lambda (I+J) Q\\
    [e_i,Q]&=-i\lambda I\Gamma_i \Gamma_+Q\quad i=1,\dots,4\\
    [e_i,Q]&=-i\lambda J\Gamma_i \Gamma_+Q\quad i=5,\dots,8\\
    [e^*_i,Q]&=-2\lambda^2 I\Gamma_i \Gamma_+Q\quad i=1,\dots,4\\
    [e^*_i,Q]&=-2\lambda^2 J\Gamma_i \Gamma_+Q\quad i=5,\dots,8\\
    [M_{ij},Q]&=\half \Gamma_{ij} Q \quad i,j=1,\dots,4~ \text{and}~
    i,j=5,\dots,8~,
  \end{aligned}
\end{equation*}
where $Q$ are the odd generators which are complex Weyl spinors.

To find the anticommutators between two odd generators of the symmetry
superalgebra, one has to compute the expression $V=\bar\varepsilon_1
\Gamma^M\varepsilon_2 \partial_M$, where
$\varepsilon_1=\varepsilon_1(\psi_1)$ and
$\varepsilon_2=\varepsilon_1(\psi_2)$ are Killing spinors. Of course,
$V$ is a Killing vector for all $\psi_1,\psi_2$. In particular for the
solution \eqref{eq:maxsol}, we find
\begin{equation*}
  \begin{aligned}[m]
    V=&-\bar\psi_1\Gamma^-\psi_2 \xi_{e_-}-\bar\psi_1\Gamma^+\psi_2
    \xi_{e_+}
    -\sum_{i=1}^8\bar\psi_1 \Gamma^i\psi_2 \xi_{e_i}\\
    & +{i\over2\lambda} \sum_{i=1}^4 \bar\psi_1 \Gamma^iI\psi_2
    \xi_{e^*_i}
    +{i\over2\lambda} \sum_{i=5}^8 \bar\psi_1 \Gamma^iJ\psi_2 \xi_{e^*_i}\\
    &+\bar\psi_1\sum_{i,j=1}^4(i\lambda \Gamma^-I)\psi_2 M_{ij}+
    \bar\psi_1\sum_{i,j=5}^8(i\lambda \Gamma^-J)\psi_2~ M_{ij} .
  \end{aligned}
\end{equation*}
{}From this expression one can read off the anticommutators
\begin{equation*}
  \begin{aligned}[m]
    \{Q,Q\}=&-\Gamma^-C^{-1}e_- - \Gamma^+C^{-1}e_+-\sum_{i=1}^8
    \Gamma^iC^{-1} e_i\\
    &+ {i\over 2\lambda} \sum_{i=1}^4 \Gamma^iIC^{-1} e_i^*
    +{i\over 2\lambda} \sum_{i=5}^8 \Gamma^iJC^{-1} e_i^* \\
    &+ i\lambda\sum_{i,j=1}^4\Gamma^-\Gamma^{ij}IC^{-1} M_{ij}+
    i\lambda\sum_{i,j=5}^8\Gamma^-\Gamma^{ij}JC^{-1} M_{ij}~.
  \end{aligned}
\end{equation*}
This concludes the computation of the structure constants of the
symmetry superalgebra $\fS$.

The solutions of the type \eqref{eq:nmaxsol} are also invariant under
the action of a superalgebra but now the action of the associated
supergroup is not transitive acting on superspace. The bosonic part of
the superalgebra is the semidirect product of the transvection Lie
algebra $\fg$ and the subalgebra of $\fso(8)$ which preserves both the
quadratic form $A$ and $F_5$. This has been explained in a similar
setting in \cite{FOPflux}. A similar computation to that presented
above reveals that the non-vanishing commutators of even and odd
generators of the superalgebra are
\begin{equation*}
  \begin{aligned}[m]
    [e_-, Q]&=i\lambda (I+J) Q\\
    [M_{ij},Q]&={1\over2} \Gamma_{ij} Q ,
  \end{aligned}
\end{equation*}
where $\Gamma_+Q=0$ and $M_{ij}$ are those generators of $\fso(8)$
that leave both the metric and $F_5$ invariant.  Then the
anticommutator of the odd generators is
\begin{equation*}
  \begin{aligned}[m]
    \{Q,Q\}=&-\Gamma^+C^{-1}e_+-\sum_{i=1}^8 \Gamma^iC^{-1} e_i\\
    &+
    {i\over 2\lambda} \sum_{i=1}^4 \Gamma^iIC^{-1} e_i^*
    +{i\over 2\lambda} \sum_{i=5}^8 \Gamma^iJC^{-1} e_i^* ~.
  \end{aligned}
\end{equation*}

\section{T-Duality, Compactification and M-Theory}

The metric and five-form field strength of the maximally
supersymmetric IIB solution that we have presented has isometries
generated by both $\partial_+ $ and $\partial _-$ and so is invariant
under translations in the space coordinate $x=x^++x^-$ and in the time
coordinate $t=x^+-x^-$.  However, the Killing spinors depend
non-trivially on all the coordinates of spacetime apart from $x^+$,
and are periodic in $x^-$ with period $2\pi/\lambda$.  They are then
periodic in both $x$ and $t$, both with the same period
$2\pi/\lambda$.  As $\partial_x$ is a Killing vector, we can
periodically identify the $x$ coordinate, $x\sim x+2\pi R$ for any $R$
to obtain a IIB supergravity solution. However, for general radii for
which $R\ne n/2\lambda$ for any integer $n$, the periodicity in $x$
will be inconsistent with the periodicity of the Killing spinors and
the compactified solution will admit no Killing spinors and preserve
no supersymmetries.  Remarkably, there is supersymmetry enhancement
for special values of the radius.  If $R= n/\lambda$ for some integer
$n$, then the solution is maximally supersymmetric with 32 Killing
spinors with periodic boundary conditions on the circle.  If $R=
(2n+1)/2\lambda$ for some integer $n$, then the solution again has 32
Killing spinors, but now with anti-periodic boundary conditions on the
circle.

For the KG solution \eqref{eq:metric},\eqref{eq:cwtype},\eqref{eq:Aij}
with $x=x^++x^-$ compactified on a spacelike circle of radius $R$, the
situation is a little more complicated, and at some special values of
the radius there are 16 Killing spinors, and at others there are 32.
From \cite{FOPflux}, for generic $R$, all supersymmetry is broken. If
$R= 4n/\mu$ for some integer $n$, then there are at least 16 periodic
Killing spinors and if $R= 2(2n+1)/\mu$, there are at least 16
anti-periodic Killing spinors.  If $R= 12n/\mu$ for some integer $n$,
then there are 32 periodic Killing spinors and if $R= 6(2n+1)/\mu$,
there are 32 anti-periodic Killing spinors.

Dimensionally reducing the IIB solution on the circle gives a solution
of 9-dimensional $N=2$ supergravity, which is an H0-brane in the
terminology of \cite{FOPflux}. As the Lie derivatives of all the
Killing spinors with respect to $\partial _x$ are non-zero, no
supersymmetries are preserved by the dimensional reduction and this
H0-brane solution of the 9-dimensional supergravity theory has no
Killing spinors.  Equivalently, the Lie algebra generator associated
with the Killing vector $\partial_x$ does not commute with the $Q$
generators, so restricting to the $x$-independent sector breaks all
supersymmetry; for further discussion on this point see for example
\cite{GGPT,FOPflux}.  Similarly, reducing the KG solution on a circle
gives a IIA solution with no Killing spinors \cite{FOPflux}.

The IIB solution on a circle can be T-dualised to obtain a solution of
the IIA theory on a dual circle of radius $\tilde R=1/R$. This is a
fundamental string solution wrapping the circle with both electric and
magnetic flux corresponding to the RR 3-form gauge field $C_3$.  This
can then be lifted to a solution of 11-dimensional supergravity which
is a membrane in a background with both electric and magnetic flux
corresponding to the 11-dimensional 3-form gauge field $C_3$.

This string or H1-brane solution of the IIA supergravity theory has no
Killing spinors that are independent of the dual coordinate $\tilde
x$, as otherwise the dimensionally reduced 9-dimensional solution
would be supersymmetric. We now turn to the question of whether this
IIA solution could have Killing spinors with non-trivial dependence on
the coordinate $\tilde x$, which is identified $\tilde x\sim \tilde
x+2\pi \tilde R$; these might arise only for special values of $\tilde
R$.  Each such Killing spinor would then lift to a Killing spinor of
the 11-dimensional solution.  The maximally supersymmetric solutions
of 11-dimensional supergravity have been classified
\cite{KG,Bonn,FOPmax} and this membrane or H2-brane is not one of
them, and so we conclude that the IIA solution cannot have 32 Killing
spinors, hence the IIA supergravity solution has strictly less
supersymmetry than the IIB supergravity solution.

However, the situation is different at the level of string theory, as
the IIB string is T-dual to the IIA string: when on a circle, the IIA
and IIB strings are the same theory, but written in terms of different
variables. The two T-dual solutions must both preserve the same amount
of supersymmetry at the level of string theory, even though they do
not at the level of the supergravity theories. This is an example of
what has been called ``supersymmetry without supersymmetry"
\cite{DLP}.  In \cite{DuffLuPopeUnt}, the $\AdS_5\times S^5$ solution
of IIB supergravity was T-dualised along the Hopf fibres of $S^5$. The
$S^5$ is a circle bundle over $\CP^2$ and dualising on the fibres
gives a IIA solution $\AdS_5\times \CP^2\times S^1$, which does not
have a spin structure.  The missing fermions arise from winding modes,
and the maximal supersymmetry of the IIB solution now becomes a IIA
fermionic symmetry that mixes ordinary field theory modes with winding
modes, and so is a non-perturbative symmetry from the world-sheet
point of view \cite{DuffLuPopeUnt}.  The situation here is similar.
The 32 supersymmetries of the IIB string solution are present in the
T-dual IIA string solution, but they are symmetries relating ordinary
field theory modes to winding modes, and these stringy supersymmetries
cannot be seen at the level of the supergravity theory.  Similarly,
the supersymmetries of the IIA solution (if any) give rise to extra
stringy supersymmetries of the IIB theory.  In the M-theory solution,
the 32 IIB supersymmetries become symmetries of M-theory that involve
membrane winding modes.

Similarly, if the time coordinate $t$ is identified with period $2\pi
T$, there will be no Killing spinors unless $T= n/2\lambda$ for
integer $n$, while at these values there will be 32 Killing spinors
which will be periodic if $n$ is even and anti-periodic if $n$ is odd.
A timelike T-duality will give a solution of the $\text{IIA}^*$ theory
\cite{Chris} which can then be lifted to the $\text{M}^*$ theory of
\cite{HullTimeLike}.

\section*{Acknowledgments}

Much of this work was done while three of us (MB,JMF,CMH) were
participating in the programme \emph{Mathematical Aspects of String
  Theory} at the Erwin Schrödinger Institute in Vienna, whom we would
like to thank for support.  The research of MB is partially supported
by EC contract CT-2000-00148.  JMF is a member of EDGE, Research
Training Network HPRN-CT-2000-00101, supported by The European Human
Potential Programme.  GP is supported by a University Research
Fellowship from the Royal Society.  This work is partially supported
by SPG grant PPA/G/S/1998/00613.  In addition, JMF would like to
acknowledge a travel grant from PPARC.

%
%

\providecommand{\bysame}{\leavevmode\hbox to3em{\hrulefill}\thinspace}

\end{document}